\documentstyle[aps,prl]{revtex}  
\draft
\begin{document}   

\def\fecha{9 January 1998}

\twocolumn[\hsize\textwidth\columnwidth\hsize
\csname @twocolumnfalse\endcsname

\title{Spacetime foam as a quantum thermal bath}
\author{Luis J. Garay}
\address{Instituto de Matem\'aticas y F\'{\i}sica Fundamental, 
CSIC, C/ Serrano 121, 28006 Madrid, Spain}
\date{\fecha}
\maketitle

\begin{abstract}

An effective model for the spacetime foam is constructed in terms
of nonlocal interactions in a classical background. In the
weak-coupling approximation, the evolution of the low-energy
density matrix is determined by a master equation that predicts 
loss of quantum coherence. Moreover, spacetime
foam can be described by a quantum thermal field that, apart from
inducing loss of coherence, gives rise to effects such as
gravitational Lamb and Stark shifts as well as quantum damping
in the evolution of the low-energy observables.
\end{abstract}

\pacs{PACS: 04.60.-m, 03.65.Bz, 04.20.Gz, 04.70.Dy
\hfill {\it gr-qc/9801024; Phys. Rev. Lett. 80, 2508 (1998)}}
 ]
\footnotetext{Copyright 1998 by The American Physical Society}

The foamlike structure of spacetime was first suggested by
Wheeler \cite{wheeler} and, since then, various components,
such as wormholes \cite{haw88,col88} and virtual black holes
\cite{96haw01}, have been proposed. The quantum fluctuations of
the geometry that constitute the spacetime foam should be of the
same order as the geometry itself at the Planck scale. This would
give rise to a minimum length \cite{95gar01} beyond which the
geometrical properties of spacetime would be lost, while on
larger scales it would look smooth and with a well-defined metric
structure.

Planck length $l_*$ might play a role analogous to the speed of
light in special relativity. In this theory, there is no physics
beyond this speed limit and its existence may be inferred through
the relativistic corrections to the Newtonian behavior. This
would mean that a quantum theory of gravity could be constructed
only on ``this side of Planck's border'' as pointed out by Markov
\cite{ma80}. In fact, the analogy between quantum gravity and
special relativity is quite close: in the latter you can
accelerate forever even though you will never reach the speed of
light; in the former, given a coordinate frame, you can reduce
the coordinate distance between two events as much as you want
even though the proper distance between them will never decrease
beyond Planck length (see Ref. \cite{95gar01}, and references
therein). This uncertainty relation $\Delta x\geq l_*$ also bears
a close resemblance to the role of $\hbar$ in quantum
mechanics: no matter which variables are used, it is not possible
to have an action $I$ smaller than $\hbar$. Indeed, the
uncertainty principle can adopt the form \cite{me91} $\Delta
I\geq \hbar$.

Spacetime foam and the related lower bound to spacetime
uncertainties would leave their imprint in low-energy physics.
Indeed, low-energy experiments would effectively suffer a
nonvanishing uncertainty coming from this lack of resolution in
spacetime measurements. Then a loss of quantum coherence would be
almost unavoidable \cite{unpred}. It could also be expected that
other effects such as transition-frequency shifts and quantum
damping, characteristic of systems in a quantum environment
\cite{91gar01}, may be present. In this Letter, we in fact show
that spacetime foam behaves as a quantum thermal bath with a
nearly Planckian temperature.

In order to build an effective theory, we will substitute the
spacetime foam, in which we possibly have 
a minimum length because the
notion of distance is not valid at such  scale, by a fixed
background with low-energy fields living on it. We will perform a
3+1 foliation of the effective spacetime that, for simplicity,
will be regarded as flat, $t$ denoting the time parameter and $x$
the spatial coordinates. The gravitational fluctuations and the
minimum length present in the original spacetime foam will be
modeled by means of nonlocal interactions that relate spacetime
points that are sufficiently close in the effective background,
where a well-defined notion of distance exists. Furthermore,
these nonlocal interactions will be described in terms of local
interactions as follows. Let $\{h_i[t]\}$ be a basis of local
gauge-invariant interactions at the spacetime point $(x,t)$ made
out of factors of the form
$l_*^{2n(1+s)-4}\left[\phi(x,t)\right]^{2n}$, $\phi$ being the
low-energy field strength of spin $s$. As a notational
convention, each index $i$ implies a dependence on the spatial
position $x$; also any contraction of indices will also entail an
integral over spatial positions. Then, we can write the nonlocal
effective interaction term in the Euclidean action as
$I_{\rm int}=\sum_N {\cal I}_N$ with
$$
{\cal I}_N=\frac{1}{N!}\int dt_1\cdots dt_N 
c^{i_1\cdots i_N}(t_1\ldots t_N)h_{i_1}[t_1]\cdots h_{i_N}[t_N].
$$
Here, $c^{i_1\cdots i_N}(t_1\ldots t_N)$ are dimensionless
functions that vanish for relative spacetime distances 
larger than the length
scale $r$ of the gravitational fluctuations. Furthermore, these
coefficients can depend only on relative positions and not on
the location of the gravitational fluctuation itself. The
physical reason for this is conservation of energy and momentum:
the fluctuations do not carry energy, momentum, or gauge charges.
Thus, diffeomorphism invariance is preserved, at least at
low-energy scales. One should not expect that at the Planck scale
this invariance still holds. However, this violation of
energy-momentum conservation is safely kept within Planck scale
limits \cite{95unr01}, where the processes will no longer be
Markovian. Furthermore, the coefficients $c^{i_1\cdots
i_N}(t_1\ldots t_N)$ will also contain a factor
$[e^{-S(r)/2}]^N$, $S(r)$ being the Euclidean action of the
gravitational fluctuation, which is of the order $(r/l_*)^2$. 

Since higher-spin $s>0$ or higher-power $n>1$ interactions are
suppressed by inverse powers of the low-energy length scale $l$,
we will concentrate on the mass term for scalar fields
$h_i[t]=l_*^{-2}\phi(x^i,t)^2$, where now the index $i$ just
keeps track of the dependence on the spatial position. A
simple calculation shows that ${\cal I}_{N}\sim\epsilon^N
(l/r)^4$, where $\epsilon=e^{-S(r)/2}(r/l_*)^4(l/l_*)^{-2}$. The
parameter $\epsilon$ has contributions from two different scales:
on the one hand, it depends on the length scale of the
gravitational fluctuations $r$ and, because of the exponential
factor, it will be very small for fluctuations of few Planck
lengths; on the other hand, it depends on the low-energy scale
$l$ through its inverse squared and will therefore be very small
far from Planck's regime. In the weak-coupling approximation,
i.e., up to second order in the expansion parameter $\epsilon$,
the trilocal and higher effective interactions do not contribute.
The terms ${\cal I}_{0}$ and ${\cal I}_{1}$ are local and can be
absorbed in the bare action (note that the cofficient $c$
appearing in ${\cal I}_{0}$ is constant and that the coefficients
$c^i(t)$ in ${\cal I}_{1}$ cannot depend on spacetime positions
because of diffeomorphism invariance). Consequently, we can write
the interaction term as a bilocal term in the Euclidean action
$$
I_{\rm int} =\frac{1}{2}\int dt dt^\prime\; c^{ij}(t-t^\prime)
h_i[t]h_j[t^\prime],
$$
where $c^{ij}(t-t^\prime)$ is of order $e^{-S(r)}$
and is concentrated within a spacetime region of size $r$.
Then, the effective partition function has the form
$Z=\int {\cal D}\phi \;e^{-I_0+I_{\rm int}}$,
$I_0$ being the bare low-energy action for the scalar field.

This bilocal effective action, when rotated back to Lorentzian
spacetime, does not lead to a unitary evolution. The reason for
this is that it is not sufficient to know the fields and their
time derivatives at an instant of time in order to know their
values at a later time: we need to know the history of the
system. 
There exist different trajectories that arrive at a given
configuration $(\phi,\dot\phi)$. The future evolution depends on
these past trajectories and not only on the values of $\phi$ and
$\dot \phi$ at that instant of time. Therefore, the system cannot
possess a well-defined Hamiltonian vector field and suffers from
an intrinsic loss of predictability \cite{89eli01}. 
This can be best dealt with by writing, up to a
determinant, the exponential of the interaction term as
\cite{96zin01} 
$$
e^{I_{int}}\sim \int {\cal D}\alpha 
e^{-\frac{1}{2} \int dt dt^\prime
\gamma_{ij}(t-t^\prime)\alpha^i(t)\alpha^j(t^\prime)} 
 e^{-\int dt \alpha^i(t) h_i[t]}.
$$
Here, the continuous matrix $\gamma_{ij}(t-t^\prime)$ is the
inverse of $c^{ij}(t-t^\prime)$, i.e., 
$\int dt^{\prime\prime}\gamma_{ik}(t-t^{\prime\prime}) 
c^{kj}(t^{\prime\prime}-t^\prime)=\delta_i^j\delta(t-t^\prime)$. 
We see that $\alpha$ is a random spacetime function subject
to a Gaussian distribution.
At second order in $\epsilon$ and lowest order in $r/l$, 
the two-point correlation function is equal to
$\langle\alpha^i(t)\alpha^j(t^\prime)\rangle=c^{ij}(t-t^\prime)$
and $\langle\alpha^i(t)\rangle=0$. Note that the Gaussian
character of the distribution for the noise $\alpha$ is
a consequence of the weak-coupling approximation (second order in
$\epsilon$), which keeps only the bilocal term in the action.
Higher-order terms would introduce deviations from this noise
distribution. The nonunitary nature of the bilocal interaction
has been encoded inside the function $\alpha$, so that, when
insisting on writing the system in terms of a Hamiltonian, an
additional sum over the part of the system that is unknown
naturally appears. Note also that we have a single function
$\alpha^i(t)$ because we are considering only one local
interaction; we will have a different function $\alpha$ for
each kind of interaction.

The Lorentzian dynamics of the low-energy field will be governed
by a master equation which can be derived after a number of steps
and approximations that are briefly outlined in what follows. For
each fixed  function $\alpha$, we first calculate the
evolution equation for the density matrix $\rho_\alpha(t)$
obtained with the Hamiltonian
$$
H_\alpha(t)= H_0[t]+ \alpha^i(t) h_i[t],
$$
$H_0[t]$ being the bare Hamiltonian of the low-energy field, and
transform this equation into the interaction picture. We then
integrate this equation between $0$ and $t$ with two iterations
and differentiate the result, so that the evolution equation
becomes an integro-differential equation for the density matrix.
Next, we perform the Gaussian average over  $\alpha$ and
expand the result up to second order in the parameter $\epsilon$,
taking into account that $\rho_\alpha$ does not depend on
$\alpha$ at zeroth order but only at first order in $\epsilon$,
i.e., $\rho_\alpha=\rho+O(\epsilon)$ with
$\rho=\langle\rho_\alpha\rangle$ (weak-coupling approximation).
We also assume that $\rho(t)$ hardly changes within a correlation
time $r$ (Markov approximation), so that $\rho(t+r)\sim \rho(t)$.
Finally, we transform the resulting equation back to the
Schr\"odinger picture. At the lowest order in $r/l$, the
result is a master equation for the low-energy density matrix
which has the form \cite{84ban01}
$$
\dot \rho= -i [ H_0,\rho]-
\int_0^\infty d\tau c^{ij}(\tau) 
\left[h_{i},\left[h_{j},\rho\right]\right].
$$
 The first term gives the Hamiltonian evolution that would
also be present in the absence of fluctuations. The second term
is a diffusion term which will be responsible for the loss of
coherence (and the subsequent 
increase of entropy). It is a direct consequence of
the foamlike structure of spacetime and the related existence of
a minimum length.

The characteristic decoherence time $\tau_{d}$ induced by the
diffusion term can be easily calculated and yields the following
ratio between the decoherence time and the low-energy length
scale: $\tau_d/l\sim e^{S(r)}(r/l_*)^{-4}$. Because of the
exponential factor, only the gravitational fluctuations whose
size is very close to Planck length will give a sufficiently
small coherence time. Slightly larger fluctuations will have a
very small effect on the unitarity of the effective theory. For
higher spins and/or powers of the field strength, the decoherence
time increases by powers of $l/l_*$. For instance, if we consider
interactions that mix fields with different spin, then the next
relevant decoherence time corresponds to the scalar-fermion
interaction term $\phi^2\bar\psi\psi$, which has an associated
decoherence ratio $\tau_d/l$ proportional to $l/l_*$. Note that
this decoherence time may be small enough for sufficiently high
energies.

Let us now go a bit further and describe spacetime foam in terms
of a quantum thermal bath. With this aim, we will consider a
system consisting of the low-energy fields coupled to a quantum bath
\cite{91gar01}. By comparing this system with the results
obtained above for gravitational fluctuations, we will see that
the latter can be substituted by a thermal bath. So, let us start
with a Hamiltonian of the form
$$
H=H_0+H_{\rm int}+H_b.
$$
$H_0$ is the bare Hamiltonian that represents the low-energy
fields and $H_b $ is the Hamiltonian of a bath that, for
simplicity, will be represented by a massless scalar field. The
interaction Hamiltonian will be of the form $H_{\rm int}=\xi^i h_i$,
where the noise operator $\xi$ is of the form 
$\xi^i(t)=i\int dk \sqrt \omega \chi(\omega)[ a^+(k)
e^{i(\omega t-k x^i)}-{\rm H.c.}]$. In this expression, $a$ and
$a^+$ are, respectively, the annihilation and creation operators associated with
the bath, $\omega=\sqrt{k^2}$, and $\chi(\omega)$ represents the
coupling between the system and the bath. This implies that
$\xi^i(t)= \chi^{ij}p_j(t)$, with $p_j(t)\equiv p(x^j,t)$ being
the momentum of the bath scalar field and 
$\chi^{ij}=\int d k \chi(\omega) \cos [k(x^i-x^j)]$
being the coupling
between the low-energy field and the bath in the position
representation.
 The coupling $\chi(\omega)$ must be such that there exists
a significant interaction with all the bath frequencies $\omega$ 
up to the
natural cuttoff $r^{-1}$. All the relevant information
about the coupling is encoded in the commutation relations and
the correlation function of the noise operator $\xi$. 

Since the commutator of the noise operator $\xi$ at different
times is a $c$ number, we can introduce the so-called commutative
noise representation \cite{91gar01}, 
which will allow us to compare this model
with that of topological fluctuations previously described. This
can be done by defining a new noise operator $\bar\alpha$ in the
following form: $\bar\alpha^i(t)\rho(t^\prime)\equiv
\frac{1}{2}[\xi^i(t),\rho(t^\prime)]_+$. It is straightforward to
check that the operator $\bar\alpha$ commutes at any time, i.e.
$[\bar\alpha^i(t),\bar\alpha^j(t^\prime)]=0$. However, this does
not mean that it commutes with everything. Indeed, the
commutator of $\bar\alpha$ with any low-energy operator $A$
is nonvanishing and has the form:
$$
[A^i(t),\bar\alpha^j(t^\prime)]=\int_{0}^t d \tau 
[A^i(t),h_k(\tau)]
\dot f^{jk} (t^\prime-\tau),
$$
where 
\begin{eqnarray}
f^{ij} (\tau)&=&\int_0^\infty 
d\omega \omega^2 G^{ij}(\omega) \cos\omega\tau,
\nonumber\\
G^{ij}(\omega)&=&\frac{\sin\omega|x^i-x^j|}{\omega|x^i-x^j|}
\chi(\omega)^2.
\nonumber
\end{eqnarray} 
The commutator above vanishes for low-energy operators that are
in the far past of the noise and is nonzero when they are in the
near past or the future. Only in the so-called first Markov
approximation the frontier among both regimes is sharply located
where both noise and low-energy fields are at the same instant of
time. Therefore, the function $f^{ij}(\tau)$ can be interpreted
as a kind of memory function.

If we assume that the bath is in a thermal state $\rho_b={\cal
Z}^{-1}e^{-H_b/T}$ with a temperature $T$ and define the average
of any operator $Q$ as $\langle Q \rangle\equiv {\rm Tr}_b(Q
\rho_b)$, we can compute the correlation function $\bar
c^{ij}(t-t^\prime) \equiv\langle \bar\alpha^i(t)
\bar\alpha^j(t^\prime)\rangle$:
$$
\bar c^{ij}(\tau)
=\int_0^\infty d\omega \omega^3 G^{ij}(\omega)[ N(\omega)+ 1/2]
\cos\omega\tau,
$$
where $N(\omega)=\left[\exp(\omega/T)-1\right]^{-1}$ is the mean
occupation number of the bath corresponding to the frequency
$\omega$. Also, it can be shown that the trace
$\langle Q \rangle$ corresponds to a Gaussian average over 
$\bar\alpha$ only in the case that the bath is in a thermal
state \cite{91gar01}, as we are considering.

We are now ready, following similar steps to those outlined
before, to write down the master equation for the low-energy
density matrix. If we keep terms only up to second order in the
expansion parameter $\bar \epsilon$ given by the product of the
thermal correlation time $\tau_{\bar\alpha}\sim 1/T$ of
$\bar\alpha$, the size of the operator $h$ and the root mean
square of $\bar\alpha$, which is of the order of
$\sqrt{\bar c}$, and we also assume that
$\tau_{\bar\alpha}/l\ll 1$, then the resulting equation has the
same form as the classical master equation obtained above with
the correlation function $c^{ij}(\tau)$ substituted for $\bar
c^{ij}(\tau)$. From the identification of both models
($\bar\alpha\equiv\alpha$), we conclude that the temperature of
the heat bath is determined by the size of the gravitational
fluctuations, i.e., $T\sim 1/r$ and that $\bar\epsilon=\epsilon\ll
1$ (weak coupling approximation). 
Note also that the coupling $\chi(\omega)$ is uniquely determined
by the correlation function $c^{ij}(\tau)$ by means of a suitable
mode expansion. The zeroth order approximation in $r/l$
that we have made in order to compare and identify both models
can be regarded as a kind of semiclassical approximation since
all the quantum features of the noise have disappeared from the
master equation.

We can however obtain a more general master equation, valid up to
second order in $\epsilon$ and with no restriction in $r/l$, that
takes into account the quantum nature of the gravitational
fluctuations. These contributions will be fairly small in the
low-energy regime, but may provide interesting information about
the higher-energy regimes in which $l$ may be of the order of a
few Planck lengths and for which the weak coupling approximation
is still valid. The quantum noise effects \cite{91gar01} are
reflected in the master equation through a term proportional to
$f^{ij}(\tau)$ and another proportional to $\bar c^{ij}(\tau)$,
both of them integrated over $\tau\in [0,\infty]$. Because of
these incomplete integrals, each term provides two different
kinds of contributions whose origin can be traced back to the
well-know formula $\int_0^\infty d\tau e^{i\omega\tau}
=\pi\delta(\omega)+P(i/\omega)$, where $P$ is the Cauchy
principal part \cite{cauchy}. Thus, the $f$ term contains a
dissipation part, necessary for the preservation of commutators,
and a contribution to what can be interpreted as a gravitational
Lamb shift. On the other hand, the $\bar c$-term gives rise to
four different contributions: The already discussed diffusion
term, another diffusion term originated from the vacuum
fluctuations of the bath and that does not vanish at zero
temperature, another contribution to the gravitational Lamb shift
and, finally, a shift in the scalar-field oscillation frequencies
that can be interpreted as a gravitational Stark effect. The size
of these effects, compared with the bare evolution, can be
calculated after some work: the thermal diffusion term is of
order $e^{-S(r)} (r/l_*)^4$, which is the only one that survived
in the previous approximations; the diffusion created by vacuum
fluctuations, the damping term, and the Stark effect are smaller
by a factor $r/l$; and the Lamb shift is smaller than the
diffusion term by a factor $(r/l)^2$.

The models described in this Letter are particularly suited to the
study of low-energy effects produced by simply connected topology
fluctuations such as virtual black holes \cite{96haw01}. Indeed,
it has been shown that virtual black holes can be represented
from the low-energy point of view by effective interactions
$h_i[t]$ like the ones employed here. The master equation can
then be interpreted as providing the evolution of the density
matrix in the presence of a bath of ubiquituous quantum
topological fluctuations of the virtual-black-hole type. 

Multiply connected fluctuations (with vanishing second Betti
number) such as wormholes \cite{haw88} can also be described as
nonlocal interactions that, in the weak-coupling approximation,
become bilocal. The coefficients $c^{ij}$ of this bilocal term do
not depend on spacetime positions since multiply connected
topology fluctuations connect spacetime points that may be far
apart from each other. Diffeomorphism invariance also requires
the spacetime independence of $c^{ij}$. This can also be seen by
analyzing these wormholes from the point of view of the universal
covering manifold, which is by definition simply connected. Here,
each wormhole is represented by two boundaries located at
infinity and suitably identified. This identification is
equivalent to introducing coefficients $c^{ij}$ that relate the
bases of the Hilbert space of wormholes in both regions of the
universal covering manifold. Since $c^{ij}$ are just the
coefficients in a change of basis, they will be constant. As a
direct consequence, the correlation time for the functions
$\alpha^i$ is infinite. This means that the functions 
$\alpha^i$ cannot be
interpreted as noise sources that are Gaussian distributed at
each spacetime point independently. Rather, they are infinitely
coherent. The Gaussian distribution to which they are subject is
therefore global, spacetime independent \cite{col88}.
Consequently, the master equation contains no diffusion term and,
actually, it predicts a unitary evolution for the density
matrix. If we still try to represent wormholes by a thermal bath
as we have done with localized gravitational fluctuations, we
soon realize that, in order to reproduce the infinite correlation
time, the couplings $\xi^i$ must be constant, that they must
commute with every other operator and, related to these two
facts, that only the zero-frequency mode of the bath can be coupled
to the low-energy fields, in agreement with the result that the
Gaussian distribution is spacetime independent and that the
effective theory is, in this case, unitary. 

Let us conclude with a brief summary. 
We have described spacetime foam in terms of
a quantum thermal field, which induces a loss of coherence in the
low-energy dynamics as well as other effects of quantum nature
such as dissipation and gravitational Lamb and Stark shifts.

I am very grateful to G.A. Mena Marug\'an, P.F.
Gonz\'alez-D\'{\i}az, C. Barcel\'o, C. Cabrillo and J.I. Cirac
for helpful discussions. I was supported by funds provided by
DGICYT and MEC (Spain) under Contract Adjunct to the Project No.
PB94--0107.


\begin{thebibliography}{99}

\bibitem{wheeler} J.A. Wheeler, in {\it Relativity, Groups and
Topology,} edited by B.S. and C.M. DeWitt (Gordon and Breach, New
York, 1964). For a recent approach, see S. Carlip, Phys. Rev.
Lett. {\bf 79}, 4071 (1997).

\bibitem{haw88} S.W. Hawking, Phys. Rev. D {\bf 37}, 904 (1988).

\bibitem{col88} S. Coleman, Nucl. Phys. {\bf B307}, 867 (1988).

\bibitem{96haw01} S.W. Hawking, Phys. Rev. D {\bf 53}, 3099 (1996).

\bibitem{95gar01} L.J. Garay, Int. J. Mod. Phys. {\bf A10}, 145
(1995).

\bibitem{ma80} M.A. Markov, Institute for  Nuclear 
Research, Moscow, Preprint P-0187, 1980 and Preprint P-0208, 1981; 
as quoted in H.-H. Borzeszkowski and H.-J. Treder, {\it The Meaning of
Quantum Gravity} (Reidel, Dordrecht, 1988).

\bibitem{me91} M.B. Mensky, Phys. Lett. {\bf A155}, 229 (1991);
Phys. Lett. {\bf A162}, 219 (1992).

\bibitem{unpred} S.W. Hawking, Commun. Math. Phys. {\bf 87}, 395
(1982).

\bibitem{91gar01} C.W. Gardiner, {\it Quantum Noise} (Springer Verlag,
Berlin, 1991).

\bibitem{95unr01} W.G. Unruh and R.M. Wald, Phys. Rev. D {\bf
52}, 2176 (1995). 

\bibitem{89eli01} D.A. Eliezer and R.P. Woodard, Nucl. Phys. {\bf
B325}, 389 (1989). 

\bibitem{96zin01} J. Zinn-Justin, {\it Quantum Field Theory and
Critical Phenomena} (Oxford University Press, Oxford, 1996), 3rd ed.

\bibitem{84ban01} T. Banks, L. Susskind, and M.E. Peskin, Nucl.
Phys. {\bf B244}, 125 (1984). 

\bibitem{cauchy} M. Reed and B. Simon, {\it Methods of Modern
Mathematical Physics I. Functional Analysis} (Academic Press, New
York, 1972).

\end{thebibliography}
\end{document}